\documentclass[amsmath,notitlepage,amssymb,aps,showkeys,floatfix,prd,a4paper,twocolumn,nofootinbib]{revtex4-1}

\usepackage{graphicx}

\usepackage{amsmath}

\usepackage{epstopdf}

\usepackage{amsfonts,amssymb}

\usepackage[utf8]{inputenc}

\usepackage{pstricks}

\usepackage{color}

\usepackage{hyperref}

\usepackage[toc]{appendix}

\usepackage[compat=1.0.0]{tikz-feynman}

\usepackage{color, colortbl}

\definecolor{Gray}{gray}{0.9}

\usepackage{float}

\usepackage{tikz-feynman}

\usepackage{forest}

\usepackage{caption}

\usepackage{subcaption}

\usepackage{verbatim}    

\usepackage{dcolumn}

\usepackage{bm}

\usepackage{epsfig}

\usepackage{braket}

\usepackage[normalem]{ulem}

\usepackage{sidecap}

\newcommand{\be}{\begin{equation}}

\newcommand{\ee}{\end{equation}}

\newcommand{\ben}{\begin{eqnarray}}

\newcommand{\een}{\end{eqnarray}}

\def\MeV{\mbox{ MeV}} 

\def\GeV{\mbox{ GeV}}

\def\keV{\mbox{ keV}} 

\def\MeV{\mbox{ MeV}} 

\def\GeV{\mbox{ GeV}} 

\def\mb{\mbox{ mb}} 

\newcommand{\pslash}{\not{\hbox{\kern-2.3pt $p$}}}

\newcommand{\pdslash}{\not{\hbox{\kern-2pt $\partial$}}}

\begin{document}

\title{Interactions of the $\chi_{c1}(4274)$ state with light mesons}

\author{A. L. M. Britto}

\email{andrebritto@ufrb.edu.br}

\affiliation{ Centro de Ciências Exatas e Tecnológicas, Universidade Federal do Recôncavo da Bahia, R. Rui Barbosa,  Cruz das Almas, 44380-000, Bahia, Brazil}

\affiliation{ Instituto de F\'isica, Universidade Federal da Bahia,Campus Universit\'ario de Ondina, 40170-115, Bahia, Brazil}

\author{L. M. Abreu}

\email{luciano.abreu@ufba.br}

\affiliation{ Instituto de F\'isica, Universidade Federal da Bahia, Campus Universit\'ario de Ondina, 40170-115, Bahia, Brazil}

\affiliation{Instituto de F\'{\i}sica, Universidade de S\~{a}o Paulo, Rua do Mat\~ao, 1371, CEP 05508-090,  S\~{a}o Paulo, SP, Brazil}

\author{F. S. Navarra}

\email{navarra@if.usp.br}

\affiliation{Instituto de F\'{\i}sica, Universidade de S\~{a}o Paulo, Rua do Mat\~ao, 1371, CEP 05508-090,  S\~{a}o Paulo, SP, Brazil}


\begin{abstract}

We investigate the interactions of the $\chi_{c1}(4274)$ state with light mesons in the hot hadron gas formed in heavy ion collisions. The vacuum and 
thermally-averaged cross sections of  production of 
$\chi_{c1}(4274)$ accompanied by light pseudoscalar and light vector mesons
as well as the corresponding inverse processes are estimated within the  
context of an effective Lagrangian approach. The results suggest non-negligible 
thermal cross-sections, with larger magnitudes for most of the suppression reactions 
than those for production. This might be a relevant feature to be considered in 
the analysis of future data  collected in heavy ion collisions.
\end{abstract}


\maketitle




\section{Introduction}

\label{Introduction}






Recently, the LHCb Collaboration reported the observation of a charmoniumlike state named $\chi_{c1}(4274)$ in the amplitude analysis of the decay $B^+\rightarrow J/ \psi \phi K^+$. Its quantum numbers have been established to be $I^G(J^{PC})=0^+(1^{++})$ with statistical significance of $6.0 \sigma$, and its measured mass and width~\cite{LHCb:2016axx,LHCb:2016nsl} 
\begin{align}
    &m =4273.3\pm 8.3^{+17.2}_{-3.6}~\MeV,
    \notag\\
    &\Gamma =56\pm 11^{8.0}_{-11}~\MeV,
\end{align}
at $5.8 \sigma$ significance. These values of mass and width are consistent with a previous measurement claimed by the CDF collaboration~\cite{CDF:2011pep}. Taking into account its isoscalar nature and decay mode, the $\chi_{c1}(4274)$ must have a minimum tetraquark content.

As a consequence, a great deal of effort has been made by the community in order to describe its properties and intrinsic quark configuration. We remark some examples of proposals for the  interpretation of the $\chi_{c1}(4274)$. Refs.~\cite{Chen:2016oma,Wang:2016dcb} considered it as a $s-$wave $cs\bar{s}\bar{c}$ tetraquark state within the framework of QCD sum rules, and ~\cite{Stancu:2009ka,Zhu:2016arf} using the compact tetraquark model. Interestingly, Ref.~\cite{Lu:2016cwr} showed that the relativized quark model proposed by Godfrey and Isgur cannot account for the compact tetraquark configuration, but for the conventional $\chi_{c1}(3^3 P_1)$ state.  On the other hand, the excited charmonium state configuration cannot be accommodated in the context of the ${}^3P_0$ model, as suggested by Ref.~\cite{Gui:2018rvv}. 

The color triplet and sextet diquark-antidiquark configuration was also employed by \cite{Agaev:2017foq}. Contrasting with experimental results, Ref.~\cite{Maiani:2016wlq} proposed that the  $\chi_{c1}(4274)$ would correspond rather to two, almost degenerate, unresolved lines with $J^{PC}=0^{++},2^{++}$. 

 From another interesting perspective, an analysis of the $\chi_{c1}(4274)$ as a $p-$wave bound state of $D_s D_{s0}(2317)$ was performed in a quasipotential Bethe-Salpeter equation approach, with a partial wave decomposition on spin parity \cite{He}. This comes from the fact that the $\chi_{c1}(4274)$ mass is just 12 MeV below the $D_s \bar{D}_{s0}(2317)$ threshold. And since the quantum numbers of $D_s $ and $\bar{D}_{s0}$  (henceforth we denote $\bar{D}_{s0}(2317)$ simply by $\bar{D}_{s0}$) are $0^{-} $ and $ 0^{+}$, the binding mechanism between these two mesons must be in $p-$wave and strong to form a bound state. However, via an effective approach Ref.~\cite{zhu2022possible} argued that the partial decay widths of the $\chi_{c1}(4274)$ do not favor the $p-$wave bound state interpretation, but indicates the possibility of a new state so-called $Y'(4274)$, which might be found in experiments such as Belle and Belle II. In the end, the intrinsic nature and the properties of the $\chi_{c1}(4274)$ is still a matter of debate, and more experimental and theoretical studies are really needed.

We believe that heavy-ion collisions (HICs) provide a promising scenario to investigate the properties of exotic states~\cite{CMS:2021znk}. As discussed in precedent works, at the end of the quark-gluon plasma phase quarks coalesce to form all types of hadronic states. The exotic states are then formed and can interact with other light hadrons  during the hadron gas phase
\cite{ChoLee1,XProd1,XProd2,Abreu:2017cof,Abreu:2018mnc,Hong:2018mpk, Abreu:2021jwm,Abreu:2022lfy,AbreuZcs}. Their final multiplicities will depend on the interaction 
cross sections, which, in turn, depend on the spatial configuration of the
quarks. Meson molecules are larger, and therefore have greater cross sections, which means that they will have a more prominent interaction with the hadronic medium than compact tetraquarks. 

In this work we investigate the interactions of the $\chi_{c1}(4274)$ state with  
light mesons  within the context of an effective Lagrangian approach. The vacuum  
and thermally-averaged cross sections of reactions involving the production of      
$\chi_{c1}(4274)$ accompanied by pseudoscalar mesons $\pi, K$ and $\eta$ and vector 
mesons $\rho, K^* $ and $\omega$ as well as the corresponding inverse processes are 
estimated.
We would like to emphasize that:

\noindent
i) The hadron gas formed in heavy ion collisions
lives for $\simeq 10$ fm and the time scale of strong interactions is 
$\simeq 1$ fm.
Therefore the multiquark states (tetraquarks or molecules) will 
inevitably interact with the light hadrons of the medium.

\noindent
ii) The careful study of the ratio $\chi_{c1}(3872) / \psi(2S)$ as a function 
of the system size published in \cite{rat-dat} made even more clear that 
the final state interactions (i.e., interactions within the hadron gas) are
crucial to understand the data \cite{xing} giving extra support to the statement i).

\noindent
iii) Simplifying assumptions concerning these interactions
such as the use of constant matrix elements or the use of geometrical arguments
to estimate the cross sections are not sufficient. The collision energies are 
of the order of the temperature, i.e., $\simeq 100 - 200$ MeV. In this energy
range the cross sections are still sensitive to resonance formation and other 
details of the interactions.

In view of these considerations, we will keep, as in previous works, 
trying to describe the multiquark interactions within the hadron gas 
with effective Lagrangians.   In the next section we will briefly describe
the formalism employed in this work. In  section III we calculate the 
interaction cross sections, in section IV we present the thermally averaged 
cross sections (called from now on simply thermal cross sections), which 
are the really relevant quantities for transport model calculations. Finally, 
in section V we present some conclusions.

\section{Formalism}

\label{Effective Lagrangian Formalism} 

\begin{figure}[h]
    \centering
\begin{subfigure}[t]{0.5\textwidth}       \centering
\begin{tikzpicture}
\begin{feynman}
\vertex (a1) {$\bar D_s (p_{1})$};
	\vertex[right=1.5cm of a1] (a2);
	\vertex[right=1.cm of a2] (a3) {$ \chi_{c1}(p_{3})$};
	\vertex[right=1.4cm of a3] (a4) {$\bar D_s (p_{1})$};
	\vertex[right=1.5cm of a4] (a5);
	\vertex[right=1.cm of a5] (a6) {$\chi_{c1} (p_{3})$};
\vertex[below=1.5cm of a1] (c1) {$D_s (p_{2})$};
\vertex[below=1.5cm of a2] (c2);

\vertex[below=1.5cm of a3] (c3) {$\pi (p_{4})$};

\vertex[below=1.5cm of a4] (c4) {$D_s (p_{2})$};

\vertex[below=1.5cm of a5] (c5);

\vertex[below=1.5cm of a6] (c6) {$\pi  (p_{4})$};

	\vertex[below=2cm of a2] (d2) {(a)};

	\vertex[below=2cm of a5] (d5) {(b)};

\diagram* {

  (a1) -- (a2), (a2) -- (a3), (c1) -- (c2), (c2) -- (c3), (c2) --

  [fermion, edge label'= $ D_{s0}$] (a2), (a4) -- (a5), (a5) -- (c6),

  (c4) -- (c5), (c5) -- (a6), (a5) -- [fermion, edge label'= $\bar D_{s0}$] (c5)

}; 

\end{feynman}

\end{tikzpicture}

\end{subfigure}\hfill

\begin{subfigure}[H]{0.5\textwidth}       \centering\begin{tikzpicture}

\begin{feynman}

\vertex (a1) {$ \bar D_{s0}(p_{1})$};

	\vertex[right=1.5cm of a1] (a2);

	\vertex[right=1.cm of a2] (a3) {$ \chi_{c1}(p_{3})$};

	\vertex[right=1.4cm of a3] (a4) {$\bar D_{s0}(p_{1})$};

	\vertex[right=1.5cm of a4] (a5);

	\vertex[right=1.cm of a5] (a6) {$ \chi_{c1}(p_{3})$};

\vertex[below=1.5cm of a1] (c1) {$D_{s0} (p_{2})$};

\vertex[below=1.5cm of a2] (c2);

\vertex[below=1.5cm of a3] (c3) {$ \pi(p_{4})$};

\vertex[below=1.5cm of a4] (c4) { $D_{s0}(p_{2})$};

\vertex[below=1.5cm of a5] (c5);

\vertex[below=1.5cm of a6] (c6) {$ \pi(p_{4})$};

	\vertex[below=2cm of a2] (d2) {(c)};

	\vertex[below=2cm of a5] (d5) {(d)};

\diagram* {

  (a1) -- (a2), (a2) -- (a3), (c1) -- (c2), (c2) -- (c3), (c2) --

  [fermion, edge label'=$  D_s$] (a2), (a4) -- (a5), (a5) -- (c6),

  (c4) -- (c5), (c5) -- (a6), (a5) -- [fermion, edge label'= $ \bar D_s$] (c5)

}; 

\end{feynman}

\end{tikzpicture}

\end{subfigure}\hfill




\begin{subfigure}[t]{0.5\textwidth}       \centering\begin{tikzpicture}

\begin{feynman}

\vertex (a1) {$\bar D_s (p_{1})$};

	\vertex[right=1.5cm of a1] (a2);

	\vertex[right=1.cm of a2] (a3) {$\chi_{c1} (p_{3})$};

	\vertex[right=1.4cm of a3] (a4) {$\bar D_s (p_{1})$};

	\vertex[right=1.5cm of a4] (a5);

	\vertex[right=1.cm of a5] (a6) {$\chi_{c1} (p_{3})$};

\vertex[below=1.5cm of a1] (c1) {$D_s (p_{2})$};

\vertex[below=1.5cm of a2] (c2);

\vertex[below=1.5cm of a3] (c3) {$\eta(p_{4})$};

\vertex[below=1.5cm of a4] (c4) {$D_s (p_{2})$};

\vertex[below=1.5cm of a5] (c5);

\vertex[below=1.5cm of a6] (c6) {$ \eta(p_{4})$};

	\vertex[below=2cm of a2] (d2) {(e)};

	\vertex[below=2cm of a5] (d5) {(f)};

\diagram* {

 (a1) -- (a2), (a2) -- (a3), (c1) -- (c2), (c2) -- (c3), (c2) --

 [fermion, edge label'= $ D_{s0}$] (a2), (a4) -- (a5), (a5) -- (c6),

 (c4) -- (c5), (c5) -- (a6), (a5) -- [fermion, edge label'= $\bar D_{s0}$] (c5)

}; 

\end{feynman}

\end{tikzpicture}

\end{subfigure}\hfill

\begin{subfigure}[t]{0.5\textwidth}       \centering\begin{tikzpicture}

\begin{feynman}

\vertex (a1) {$\bar D_{s0}(p_{1})$};

	\vertex[right=1.5cm of a1] (a2);

	\vertex[right=1.cm of a2] (a3) {$\chi_{c1} (p_{3})$};

	\vertex[right=1.4cm of a3] (a4) {$\bar D_{s0} (p_{1})$};

	\vertex[right=1.5cm of a4] (a5);

	\vertex[right=1.cm of a5] (a6) {$\chi_{c1}(p_{3})$};

\vertex[below=1.5cm of a1] (c1) {$ D_{s0}(p_{2})$};

\vertex[below=1.5cm of a2] (c2);

\vertex[below=1.5cm of a3] (c3) {$\eta (p_{4})$};

\vertex[below=1.5cm of a4] (c4) {$D_{s0} (p_{2})$};

\vertex[below=1.5cm of a5] (c5);

\vertex[below=1.5cm of a6] (c6) {$ \eta(p_{4})$};

	\vertex[below=2cm of a2] (d2) {(g)};

	\vertex[below=2cm of a5] (d5) {(h)};

\diagram* {

 (a1) -- (a2), (a2) -- (a3), (c1) -- (c2), (c2) -- (c3), (c2) --

 [fermion, edge label'= $ D_s$] (a2), (a4) -- (a5), (a5) -- (c6),

 (c4) -- (c5), (c5) -- (a6), (a5) -- [fermion, edge label'= $\bar D_s$] (c5)

}; 

\end{feynman}

\end{tikzpicture}

\end{subfigure}



\begin{subfigure}[t]{0.5\textwidth}       \centering\begin{tikzpicture}

\begin{feynman}

\vertex (a1) {$\bar  D_{s0}(p_{1})$};

	\vertex[right=1.5cm of a1] (a2);

	\vertex[right=1.cm of a2] (a3) {$\chi_{c1} (p_{3})$};

	\vertex[right=1.4cm of a3] (a4) {$ D_{s0}(p_{1})$};

	\vertex[right=1.5cm of a4] (a5);

	\vertex[right=1.cm of a5] (a6) {$ \chi_{c1}(p_{3})$};

\vertex[below=1.5cm of a1] (c1) {$ D_s^*(p_{2})$};

\vertex[below=1.5cm of a2] (c2);

\vertex[below=1.5cm of a3] (c3) {$ \eta(p_{4})$};

\vertex[below=1.5cm of a4] (c4) {$\bar  D^*(p_{2})$};

\vertex[below=1.5cm of a5] (c5);

\vertex[below=1.5cm of a6] (c6) {$ K(p_{4})$};

	\vertex[below=2cm of a2] (d2) {(i)};

	\vertex[below=2cm of a5] (d5) {(j)};

\diagram* {

  (a1) -- (a2), (a2) -- (a3), (c1) -- (c2), (c2) -- (c3), (c2) --

  [fermion, edge label'= $  D_s$] (a2), (a4) -- (a5), (a5) -- (a6),

  (c4) -- (c5), (c5) -- (c6), (c5) -- [fermion, edge label'= $ \bar D_s$] (a5)

}; 

\end{feynman}

\end{tikzpicture}

\end{subfigure}\hfill

\begin{subfigure}[t]{0.5\textwidth}       \centering\begin{tikzpicture}

\begin{feynman}

   \vertex (b) ;

    \vertex [left=1.cm of b] (a) {$  D_s (p_{1})$};


    \vertex [ right=1.cm of b] (f1) {$\chi_{c1}(p_3)$};

    \vertex [below =1.5cm of b] (c);

    \vertex [ right=1.cm of c] (f2) {$ K(p_{4})$};

    \vertex [left=1.cm of c] (f3) {$\bar D (p_{2})$};

    \vertex[below=2cm of b] (d2) {(k)};

    \diagram* {

      (a) -- (b) -- (f1),

      (c) -- [fermion, edge label'= $ \bar D_{s0}$] (b),

      (c) --  (f2),

      (c) -- (f3),    };

\end{feynman}

\end{tikzpicture}

\end{subfigure}

\caption{Born diagrams describing the production of the 
$\chi_{c1}(4274)$ (denoted by $\chi_{c1}$) and a light 
pseudoscalar meson (without specification of 
the charges of the particles).}
\label{DIAG1}
\end{figure}
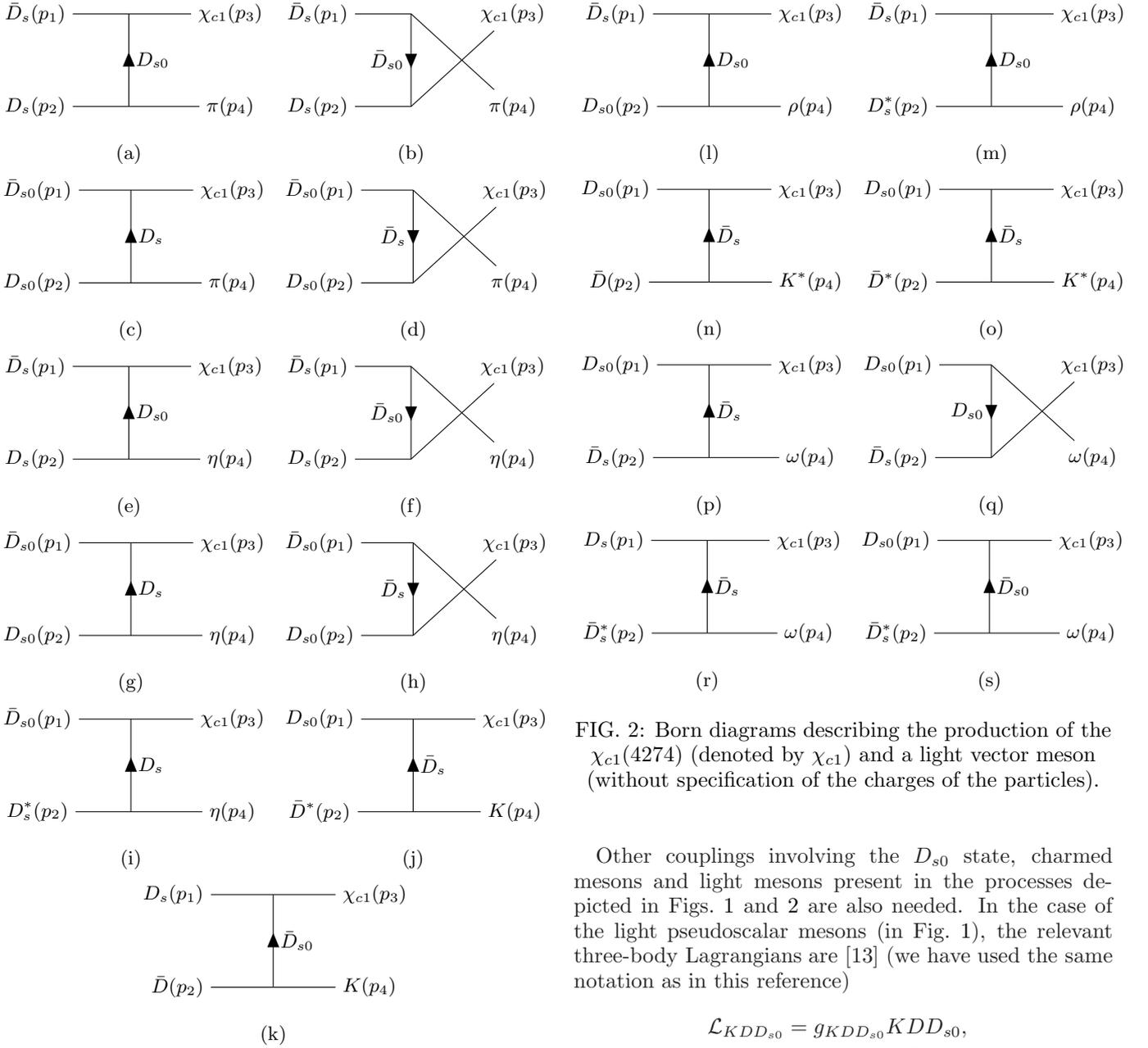
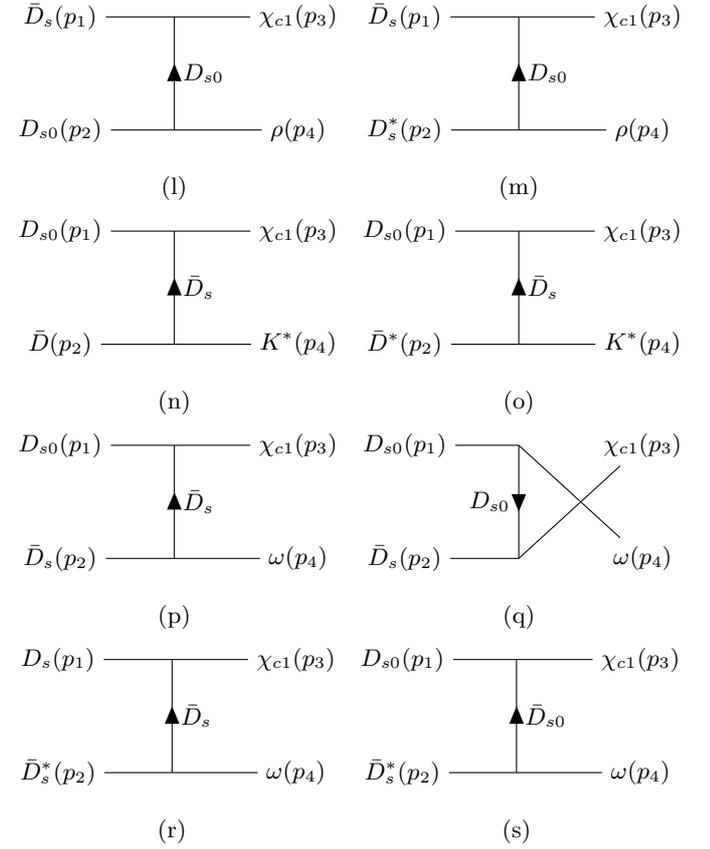
\begin{figure}[h]
\begin{subfigure}[t]{0.5\textwidth}\centering

\begin{tikzpicture}

\begin{feynman}

\vertex (a1) {$\bar D_s(p_{1})$};

	\vertex[right=1.5cm of a1] (a2);

	\vertex[right=1.cm of a2] (a3) {$ \chi_{c1} (p_{3})$};

	\vertex[right=1.4cm of a3] (a4) {$\bar  D_s(p_{1})$};

	\vertex[right=1.5cm of a4] (a5);

	\vertex[right=1.cm of a5] (a6) {$ \chi_{c1}(p_{3})$};

\vertex[below=1.5cm of a1] (c1) {$D_{s0} (p_{2})$};

\vertex[below=1.5cm of a2] (c2);

\vertex[below=1.5cm of a3] (c3) {$ \rho(p_{4})$};

\vertex[below=1.5cm of a4] (c4) {$ D_s^*(p_{2})$};

\vertex[below=1.5cm of a5] (c5);

\vertex[below=1.5cm of a6] (c6) {$ \rho(p_{4})$};

	\vertex[below=2cm of a2] (d2) {(l)};

	\vertex[below=2cm of a5] (d5) {(m)};

\diagram* {(a4) -- (a5), (a5) -- (a6),

  (c4) -- (c5), (c5) -- (c6), (c5) -- [fermion, edge label'= $ D_{s0}$] (a5),

  (a1) -- (a2), (a2) -- (a3), (c1) -- (c2), (c2) -- (c3), (c2) -- 

  [fermion, edge label'= $ D_{s0}$] (a2) 

}; 

\end{feynman}

\end{tikzpicture}

\end{subfigure}\hfill




\begin{subfigure}[t]{0.5\textwidth}\centering

\begin{tikzpicture}

\begin{feynman}

\vertex (a1) {$ D_{s0}(p_{1})$};

	\vertex[right=1.5cm of a1] (a2);

	\vertex[right=1.cm of a2] (a3) {$ \chi_{c1}(p_{3})$};

	\vertex[right=1.4cm of a3] (a4) {$D_{s0} (p_{1})$};

	\vertex[right=1.5cm of a4] (a5);

	\vertex[right=1.cm of a5] (a6) {$ \chi_{c1}(p_{3})$};

\vertex[below=1.5cm of a1] (c1) {$ \bar{D}(p_{2})$};

\vertex[below=1.5cm of a2] (c2);

\vertex[below=1.5cm of a3] (c3) {$ K^{*}(p_{4})$};

\vertex[below=1.5cm of a4] (c4) {$ \bar{D}^*(p_{2})$};

\vertex[below=1.5cm of a5] (c5);

\vertex[below=1.5cm of a6] (c6) {$ K^{*}(p_{4})$};

	\vertex[below=2cm of a2] (d2) {(n)};

	\vertex[below=2cm of a5] (d5) {(o)};

\diagram* {(a4) -- (a5), (a5) -- (a6),

  (c4) -- (c5), (c5) -- (c6), (c5) -- [fermion, edge label'= $\bar D_s$] (a5),

  (a1) -- (a2), (a2) -- (a3), (c1) -- (c2), (c2) -- (c3), (c2) -- 

  [fermion, edge label'= $\bar D_s$] (a2) 

}; 

\end{feynman}

\end{tikzpicture}

\end{subfigure}\hfill

\begin{subfigure}[b]{0.5\textwidth}

\centering

\begin{tikzpicture}

\begin{feynman}

\vertex (a1) {$D_{s0} (p_{1})$};

	\vertex[right=1.5cm of a1] (a2);

	\vertex[right=1.cm of a2] (a3) {$ \chi_{c1}(p_{3})$};

	\vertex[right=1.4cm of a3] (a4) {$ D_{s0}(p_{1})$};

	\vertex[right=1.5cm of a4] (a5);

	\vertex[right=1.cm of a5] (a6) {$ \chi_{c1}(p_{3})$};

\vertex[below=1.5cm of a1] (c1) {$\bar{D}_{s} (p_{2})$};

\vertex[below=1.5cm of a2] (c2);

\vertex[below=1.5cm of a3] (c3) {$ \omega(p_{4})$};

\vertex[below=1.5cm of a4] (c4) {$ \bar{D}_s(p_{2})$};

\vertex[below=1.5cm of a5] (c5);

\vertex[below=1.5cm of a6] (c6) {$ \omega(p_{4})$};

	\vertex[below=2cm of a2] (d2) {(p)};

	\vertex[below=2cm of a5] (d5) {(q)};

\diagram* {

  (a1) -- (a2), (a2) -- (a3), (c1) -- (c2), (c2) -- (c3), (c2) --

  [fermion, edge label'= $\bar D_s$] (a2),(a4) -- (a5), (a5) -- (c6),

  (c4) -- (c5), (c5) -- (a6), (a5) -- [fermion, edge label'= $D_{s0}$] (c5)

}; 

\end{feynman}

\end{tikzpicture}

\end{subfigure}\hfill

\begin{subfigure}[b]{0.5\textwidth}

\centering

\begin{tikzpicture}

  \begin{feynman}

\vertex (a1) {$D_s(p_{1})$};

	\vertex[right=1.5cm of a1] (a2);

	\vertex[right=1.cm of a2] (a3) {$ \chi_{c1} (p_{3})$};

	\vertex[right=1.4cm of a3] (a4) {$D_{s0}(p_1)$};

	\vertex[right=1.5cm of a4] (a5);

	\vertex[right=1.cm of a5] (a6) {$ \chi_{c1}(p_{3})$};

\vertex[below=1.5cm of a1] (c1) {$\bar  D_s^*(p_{2})$};

\vertex[below=1.5cm of a2] (c2);

\vertex[below=1.5cm of a3] (c3) {$\omega (p_{4})$};

\vertex[below=1.5cm of a4] (c4) {$\bar D_s^{*}(p_2) $};

\vertex[below=1.5cm of a5] (c5);

\vertex[below=1.5cm of a6] (c6) {$\omega (p_{4})$};

	\vertex[below=2cm of a2] (d2) {(r)};

	\vertex[below=2cm of a5] (d5) {(s)};

\diagram* {(a4) -- (a5), (a5) -- (a6),

  (c4) -- (c5), (c5) -- (c6), (c5) -- [fermion, edge label'= $\bar D_{s0}$] (a5),

  (a1) -- (a2), (a2) -- (a3), (c1) -- (c2), (c2) -- (c3), (c2) -- 

  [fermion, edge label'= $\bar D_s$] (a2) 

};

\end{feynman}

\end{tikzpicture}\end{subfigure}

\caption{Born diagrams describing the production of the $\chi_{c1}(4274)$ (denoted by $\chi_{c1}$) and 
a light vector meson (without specification of the charges of the particles).}       
\label{DIAG2}
\end{figure}

In this section, we will investigate the interactions of the $\chi_{c1}(4274)$
state with the lightest pseudoscalar mesons ($\pi, K$ and $\eta$) and with the 
lightest vector mesons ($\rho$, $\omega$ and $K^*$). We will study the reactions 
$  \chi_{c1}(4274) \, + \, (\pi, K , \eta, \rho, K^* ,\omega)     
\to  (\bar{D}_{s}^{} D^{(*)} , D_{s0} D_s^{(*)}, D_{s0}  D_{s0} )$, as well as 
the inverse processes. 
The lowest-order Born diagrams that contribute to the processes of our interest 
are shown in Figs.~\ref{DIAG1} and~\ref{DIAG2}. To calculate their respective 
amplitudes, we use an effective theory formalism in which the vector mesons are 
interpreted as dynamical gauge bosons of the hidden $U(N)_V$ local 
symmetry~(see Refs.~\cite{XProd1,XProd2,Abreu:2017cof,Abreu:2018mnc} for a more
detailed discussion). In particular, the following effective Lagrangians involving 
the light and charmed mesons are employed:
\begin{eqnarray}
\mathcal{L}_{PPV} & = & -ig_{PPV}\langle V^\mu[P,\partial_\mu P]\rangle , 
    \label{LPPV} \\
\mathcal{L}_{VVP} & = & \frac{g_{VVP}}{\sqrt{2}}
\varepsilon^{\mu\nu\alpha\beta}\langle
\partial_\mu V_\nu \partial_\alpha V_\beta 
\rangle, 
\label{LVVP}
\end{eqnarray}
where $P$ and $V$ are the matrices in $\rm SU(4)$ flavor space containing 
the pseudoscalar and vector meson fields,
\begin{widetext}
\begin{eqnarray}\raggedleft
 P & = &    \left(
\begin{array}{cccc}
 \frac{\eta }{\sqrt{6}}+\frac{{\eta_c}}{\sqrt{12}}
+\frac{{\pi^0}}{\sqrt{2}} & {\pi^+} & {K^+} & {\bar{D}^0} \\
 {\pi^-} & \frac{\eta }{\sqrt{6}}+\frac{{\eta_c}}{\sqrt{12}}
-\frac{{\pi^0}}{\sqrt{2}} & {K^0} & {D^-} \\
 {K^-} & {\bar{K}^0} & \frac{{\eta_c}}{\sqrt{12}}-\frac{2 \eta }{\sqrt{6}} & 
{D_s^-} \\
 {D^0} & {D^+} & {D_s^+} & {\eta_c} \\
\end{array}
\right),\\
V & = & \left(
\begin{array}{cccc}
 \frac{\rho^0}{\sqrt{2}}+\frac{{\omega }}{\sqrt{6}}+\frac{J/\psi}{\sqrt{12}} & 
{{\rho^+}} & {{K^{*+}}} & {{\bar{D}^{*0}}} \\
 {{\rho^-}} & -\frac{\rho^0}{\sqrt{2}}+\frac{{\omega }}{\sqrt{6}}+
\frac{J/\psi}{\sqrt{12}}  & {{K^{*0}}} & {{D^{*-}}}
   \\
 {{K^{*-}}} & {{\bar{K}^{*0}}} & -2\frac{{\omega }}{\sqrt{6}}+
\frac{J/\psi}{\sqrt{12}}  & {{D_s^{*-}}} \\
 {{D^{*0}}} & {{D^{*+}}} & {{D_s^{*+}}} & -3\frac{J/\psi}{\sqrt{12}} \\
\end{array}
\right) ; 
\label{matrix}
\end{eqnarray}
\end{widetext}
the coupling constants are given by:              
\begin{eqnarray}
g_{PPV} & = & \frac{m_V}{2f_\pi} \frac{m_{D^*}}{m_{K^*}}, \nonumber   \\
g_{VVP} &  =  &\frac{3m^2_V}{16\pi^2f^3_\pi}, 
\label{eq:CouplingConstants2}
\end{eqnarray}
with $m_V$ being the mass of the vector meson; we take it as the mass of
the $\rho$ meson and $f_\pi$ is the pion decay constant. As pointed in      
Ref.~\cite{XProd1}, the factor $m_{D^*}/m_{K^*} $ in the coupling $g_{PPV}$ 
is introduced in order to reproduce the experimental decay width found for
the process $D^* \to D\pi$, and comes from heavy-quark symmetry considerations. 


Let us now introduce the vertex associated to the $\chi_{c1}(4274)$ state.  
The quantum numbers of  $\chi_{c1}(4274)$  are $I(J^{PC})=0(1^{++})$ and those of  
$D_s$ and $D_{s0}$ are $J^{P}=0^-$ and $J^P=0^+$ respectively. Hence, for parity 
reasons, the vertex $ \chi_{c1} \, D_s \,D_{s0}$ must be in relative $P$-wave. This can be
implemented by a term with a derivative coupling in the Lagrangian. 
The appropriate  effective Lagrangian is then given by ~\cite{zhu2022possible}: 
\begin{align}
\mathcal{L}_{\chi_{c1}}=\frac{1}{\sqrt{2}}g_{\chi_{c1}D_s\bar{D}_{s0}}  \chi_{c1}^{\mu}\left[ D^+_s  
\overset{\leftrightarrow}{\partial}_\mu D^{-}_{s0}- D^-_s 
\overset{\leftrightarrow}{\partial}_\mu D^+_{s0}
    \right] ,  \label{lagranY}
\end{align}
where  $ \chi_{c1}^{\mu}$ stands for the field associated to $\chi_{c1}(4274)$ state;  
$g_{\chi_{c1}D_s\bar{D}_{s0}}$ is the effective coupling constant taken from the  
analysis of the $\chi_{c1}(4274)$ width performed in Ref.~\cite{zhu2022possible}: 
$g_{\chi_{c1}D_s\bar{D}_{s0}} = 13.34^{+1.11}_{-0.89}$. 

Other couplings involving the $D_{s0}$ state, charmed mesons and light mesons
present in the processes depicted in Figs.~\ref{DIAG1} and \ref{DIAG2} are also  
needed. In the case of the light pseudoscalar mesons (in Fig.~\ref{DIAG1}), the 
relevant three-body Lagrangians are~\cite{zhu2022possible}
(we have used the same notation as in this reference)
\begin{align}
\mathcal{L}_{KDD_{s0}}&=g_{KDD_{s0}}KD D_{s0},\notag\\
\mathcal{L}_{\pi^0D_sD_{s0}}&=g_{\pi^0D_sD_{s0}}\pi^0  D_s D_{s0} , \nonumber \\
\mathcal{L}_{\eta D_sD_{s0}}&=g_{\eta D_sD_{s0}}\eta D_s D_{s0},
   \label{ds0vertex}
\end{align}
where the coupling constants take the values 
$g_{KDD_{s0}}=10.21\GeV$, $g_{\pi^0D_sD_{s0}}=1.3124 \GeV$ and 
$g_{\eta D_sD_{s0}}=6.40\GeV$. 
For the three-body vertices involving the $D_{s0}$ state, charmed mesons and light  
vector mesons (in Fig.~\ref{DIAG2}), we employ the following Lagrangians
~\cite{VMD}
    \begin{align}
          \mathcal{L}_{VD_{s0}D_{s0}} &= -ig_{V D_ {s0}D_ {s0}}V^\mu 
D_{s0}^-\overset{\leftrightarrow}{\partial}_\mu D_{s0}^+ , 
\notag\\
  \mathcal{L}_{V D_{s0}D_s^*}&=g_{V D_{s0}D_s^*}[D_ {s0}^- D_ {s\mu\nu}^{*+}
- D_ {s0}^+D_{s\mu\nu}^{*-}]V_{\mu\nu},
\label{VMDvertex}
    \end{align}
where $V_{\mu\nu}=\partial_\mu V_\nu-\partial_\nu V_\mu$. The coupling constants 
$g_{V D_ {s0}D_ {s0}} $ and $ g_{V D_{s0}D_s^*} $ are estimated through the vector  
dominance (VMD) model~\cite{VMD}. Accordingly, the virtual photon in the   
$D_{s0}^+e^-\rightarrow D_{s0}^+e^-$ scattering is coupled to the vector meson via 
the photon-vector-meson mixing term
\begin{align}
\gamma_V V^\mu A_\mu,
\end{align}
where $A_\mu$ is the photon field; $\gamma_V$ is the photon-vector-meson       
mixing amplitude, determined from the width of the vector-to-electron-positron 
decay,
\begin{align}
\Gamma_V = \frac{\alpha_{em}\gamma_V}{3m_V^3},
\end{align}
where $\alpha_{em}$ is the electromagnetic fine-structure constant and $m_V$ is     
the vector mass. For the $\omega$ and $\rho$ mesons we have $\Gamma_\omega=0.6\keV$  
and $\Gamma_{\rho}=6.96\keV$ respectively \cite{PDG}, which yields the mixing  
amplitudes $\gamma_\omega=0.011\GeV^2$ and $\gamma_{\rho}=0.037\GeV^2$. Next,      
the coupling constants $g_{\omega D_ {s0}D_ {s0}}$ and $g_{\rho^0 D_ {s0}D_ {s0}}$ 
can be estimated with the following expression: 
\begin{equation}
        \frac{\gamma_V g_{V D_ {s0}D_ {s0}} }{m_V^2}=\frac{1}{3}e; 
        \label{coupl1}
\end{equation}
which gives $g_{\omega D_ {s0}D_ {s0}}=5.50$ and $g_{\rho^0 D_ {s0}D_ {s0}}=1.66$. 
To obtain $g_{V D_{s0}D_s^*}$, we make use of an extension of Eq.~(\ref{coupl1}), 
i.e.  
\begin{equation}
      \frac{\gamma_V g_{V D_{s0}D_s^*} }{m_V^2}=\frac{1}{3}e g_{\gamma D_{s0}D_s^*}, 
      \label{coupl2}
\end{equation}
where $ g_{\gamma D_{s0}D_s^*}$ is the coupling constant of the vertex involving 
the photon, $D_{s0} $ and $ D^{*}_s$, which from theoretical estimates is          
$g_{\gamma D_{s0}D^{*}_s} \geq 3.02\times 10^{-2}\GeV$ \cite{VMD,Faessler:2007gv}, 
Therefore, we can write  
\begin{align}
        g_{V D_{s0}D_s^*}=g_{V D_ {s0} D_ {s0}} \times g_{\gamma D_{s0}D_s^*}.
\end{align}    
Taking the smallest value of $g_{\gamma D_{s0}D_s^*}$, we obtain 
$g_{\omega D_{s0}D_s^*}=0.17 \GeV$ and $g_{\rho D_{s0}D_s^*}=0.050\GeV$.        
Then, the effective model above allows us to write the amplitudes corresponding to 
the diagrams depicted in Figs.~\ref{DIAG1} and~\ref{DIAG2} as:
\begin{align}
M_{D_sD_s\rightarrow \chi_{c1}\pi}=&M^{(a)}+M^{(b)},\notag\\
M_{D_{s0}D_{s0}\rightarrow \chi_{c1}\pi }=&M^{(c)}+M^{(d)},\notag\\
M_{ D_sD_s\rightarrow \chi_{c1}\eta}=&M^{(e)}+M^{(f)},\notag\\
M_{D_{s0}D_{s0}\rightarrow \chi_{c1}\eta}=&M^{(g)}+M^{(h)},\notag\\
M_{ D_{s0}D_s^*\rightarrow \chi_{c1}\eta}=&M^{(i)},\notag\\
M_{ D_{s0}D^*\rightarrow \chi_{c1}K}=&M^{(j)},\notag\\
M_{D_sD\rightarrow \chi_{c1}K\rightarrow  }=&M^{(k)},\notag\\
M_{D_sD_{s0}\rightarrow \chi_{c1}\rho}=&M^{(l)},\notag\\
M_{D_sD_s^*\rightarrow  \chi_{c1}\rho}=&M^{(m)},\notag\\
M_{D_{s0}D\rightarrow  \chi_{c1}K^{*}}=&M^{(n)},\notag\\
M_{D_{s0}D^*\rightarrow \chi_{c1}K^{*}}=&M^{(o)},\notag\\
M_{ D_{s0}D_s\rightarrow  \chi_{c1}\omega}=&M^{(p)}+M^{(q)},\notag\\
M_{ D_sD_s^*\rightarrow  \chi_{c1}\omega}=&M^{(r)},\notag\\
M_{D_{s0}D_s^*\rightarrow  \chi_{c1}\omega}=&M^{(s)},
\label{ampl1}
\end{align}
The explicit expressions are described in the Appendix~\ref{appendice}. 

\section{Cross Sections}
 We define the isospin-spin-averaged cross section in the center of mass (CM) 
frame for a given reaction $ab\rightarrow cd$ in Eq~(\ref{ampl1}) as:
\begin{align}
\sigma_{ab\rightarrow cd}= \frac{1}{64\pi^2 s}
\frac{|\vec{p}_{cd}|}{|\vec{p}_{ab}|}\frac{1}{g_a g_b}\int d\Omega
\sum_{S,I} |M_{ab\rightarrow cd}|^2,
\end{align}
where $g_{a,b}=(2I_{a,b}+1)(2S_{a,b}+1)$ is the degeneracy factor of the of the  
particles in initial state; $s$ is the squared center-of-mass energy;         
$| \vec{p}_{ab} |$ and $| \vec{p}_{cd} |$ are the moduli of the three-momenta  
in CM frame of the initial and final particles, respectively. The summation is 
performed over the spin and isospin of the initial and final states, with the latter  
being rewritten in terms of the particle basis with the explicit charges of the 
particles in the initial state, i.e.
\begin{equation}
 \displaystyle  \sum_{I} |M_{ab\rightarrow cd}|^2\rightarrow\sum_{Q_a,Q_b} 
|M_{ab\rightarrow cd}^{(Q_1,Q_2)}|^2.
\end{equation}

The cross sections of the corresponding inverse reactions are evaluated by means 
of the detailed balance relation, 
\begin{align}
g_a g_b|\vec{p}_{ab}|^2\sigma_{ab\rightarrow cd}
=g_c g_d|\vec{p}_{cd}|^2\sigma_{cd\rightarrow ab}.
\label{detbal}
\end{align}
Furthermore, to take into account the finite size of the hadrons and to suppress  
the artificial growth of the cross sections at large momenta, we make use of a 
monopole-like form factor,
\begin{equation}
F(\vec{q})=\frac{\Lambda^2}{\Lambda^2+\vec{q}^2},
\end{equation}
where $\vec{q}$ is the transferred three-momentum in $t(u)$ channel,  and      
$\Lambda$ is the cutoff, which we choose to be $\Lambda = 2.0 $ GeV. 
This type of form factor has been extensively employed in literature, and a 
detailed discussion on its  role is found in Ref.~\cite{Abreu:2021jwm}.

The calculations are done with the isospin-averaged masses reported in the      
PDG~\cite{PDG}. Besides, to take into account the uncertainties in the coupling  
constant $g_{\chi_{c1}D_s D_{s0}}$, the results are presented in terms of bands associated 
to the smallest and largest possible values of $g_{\chi_{c1}D_s D_{s0}}$.

The cross sections for the processes discussed in previous section are plotted     
in Figs.~\ref{YpScalar} and~\ref{Yvector} as functions of the CM energy $\sqrt{s}$, 
as well as those for the corresponding inverse reactions. In the case of 
$\chi_{c1}(4274)$ 
production, all cross sections are endothermic, showing a fast growth in the region  
very close to the threshold, with exception of the channel $ D_{s0}D_{s0} \to \chi_{c1}\pi $. 
Considering the region up to $400 \MeV$ above the corresponding thresholds, the  
different channels present a wide range of magnitudes     
$( \sim  10^{-6} -  10^{0} \mb$. 
In particular, for the $\chi_{c1}(4274)$ production accompanied by pion and 
$\eta $ mesons, 
the channel with initial state $D_s^0 \bar{D}_s ^{0}$ yields the dominant 
contributions, while the other cases are at least one order of magnitude smaller.  
We also remark that the channels involving the kaon or $K^*$ mesons        
give contributions of similar order. Besides, the reactions involving the   
$\rho^0$ mesons with the initial or final state $ D_{s}\bar{D}_s^*$ present  
the smallest cross sections due to the smaller coupling constant of the 
$\rho D_{s0}D_s^*$ interaction.

\begin{widetext}
\onecolumngrid
\begin{figure}[h]
\centering
\includegraphics[{width=0.32\linewidth}]{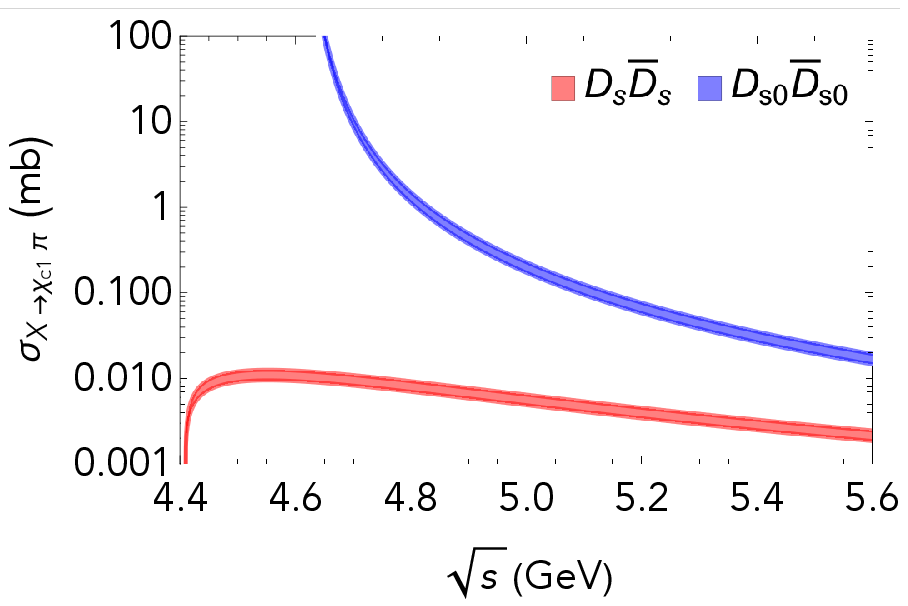}
    \includegraphics[{width=0.32\linewidth}]{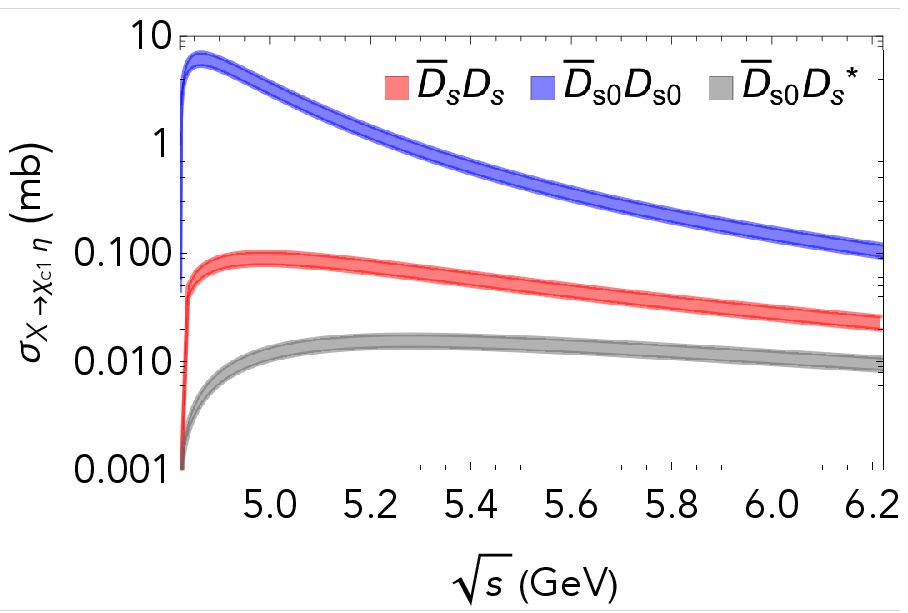} 
    \includegraphics[{width=0.32\linewidth}]{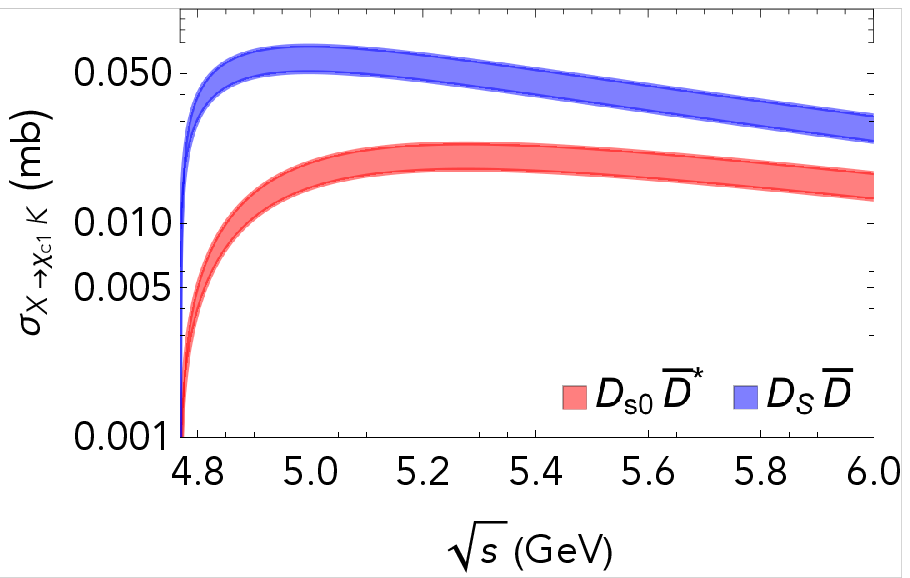}
\\
    \includegraphics[{width=0.32\linewidth}]{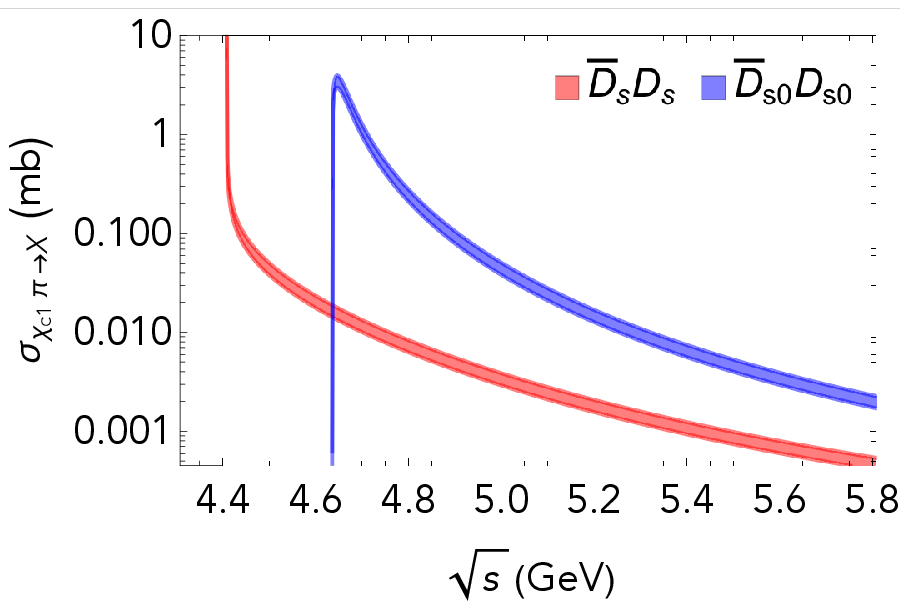}
    \includegraphics[{width=0.32\linewidth}]{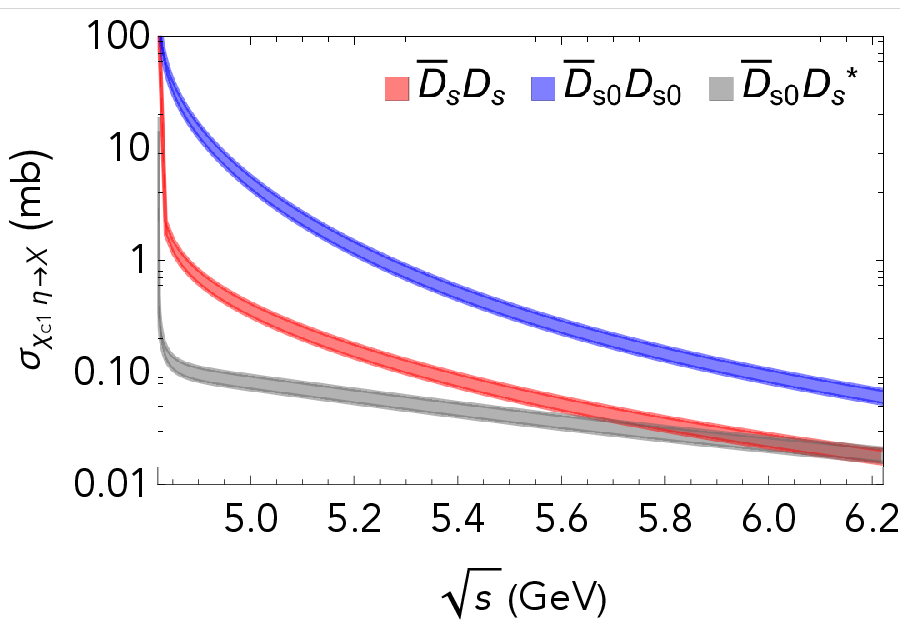}
%
    \includegraphics[{width=0.32\linewidth}]{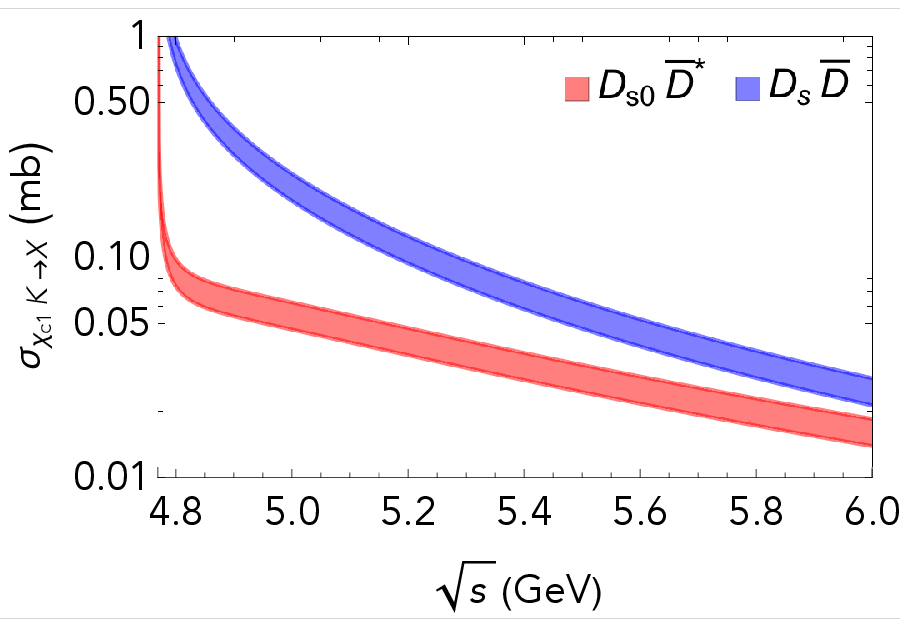}
\caption{Top: cross sections for the production processes 
$ \chi_{c1}(4274) \pi  $ (left),
$ \chi_{c1}(4274) \eta$ (center) and $ \chi_{c1}(4274) K$ (right), as functions of          
$\sqrt{s}$. Bottom: cross sections for the corresponding inverse reactions.}
\label{YpScalar} 
\end{figure}
\end{widetext}
\twocolumngrid
Let us now look at the $\chi_{c1}(4274)$-suppression processes in 
Figs.~\ref{YpScalar}  
and~\ref{Yvector}. As expected only the process $\chi_{c1}\pi\rightarrow D_{s0}D_{s0}$  
is endothermic; the other absorption cross sections are exothermic, becoming     
very large at the threshold. Above the threshold, these cross sections have very 
distinct magnitudes. Most importantly, when we compare $\chi_{c1}(4274)$  
absorption and 
production by comoving light mesons in the relevant region of energies for heavy  
ion collisions ($\sqrt{s} -\sqrt{s_0} < \, 0.6$ GeV), in general the absorption 
cross sections are greater than the  production ones. This feature reflects the  
differences of these reactions concerning the phase space as well the degeneracy 
factors encoded in Eq.~(\ref{detbal}).
%
\begin{widetext}
\onecolumngrid
\begin{figure}[h]
    \includegraphics[{width=0.32\linewidth}]{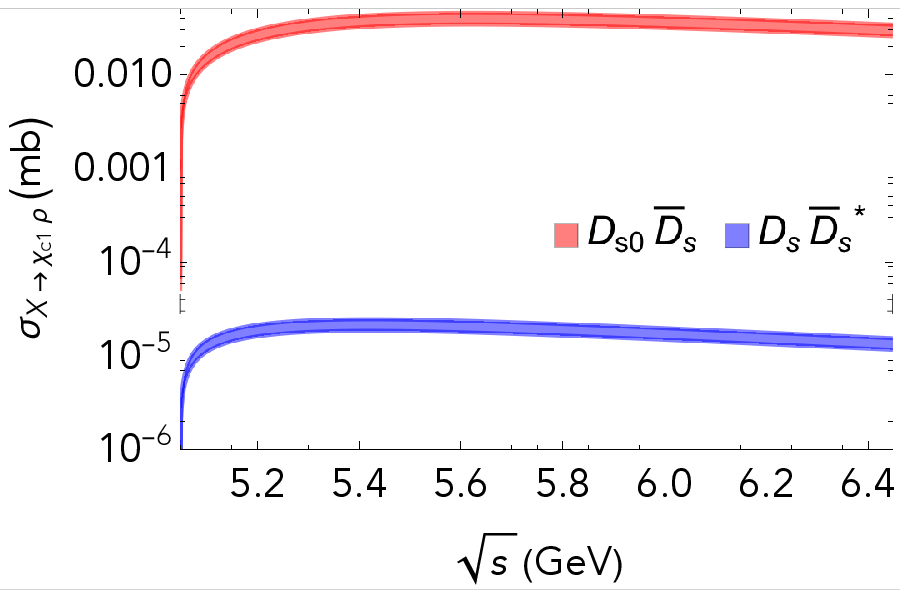}
    \includegraphics[{width=0.32\linewidth}]{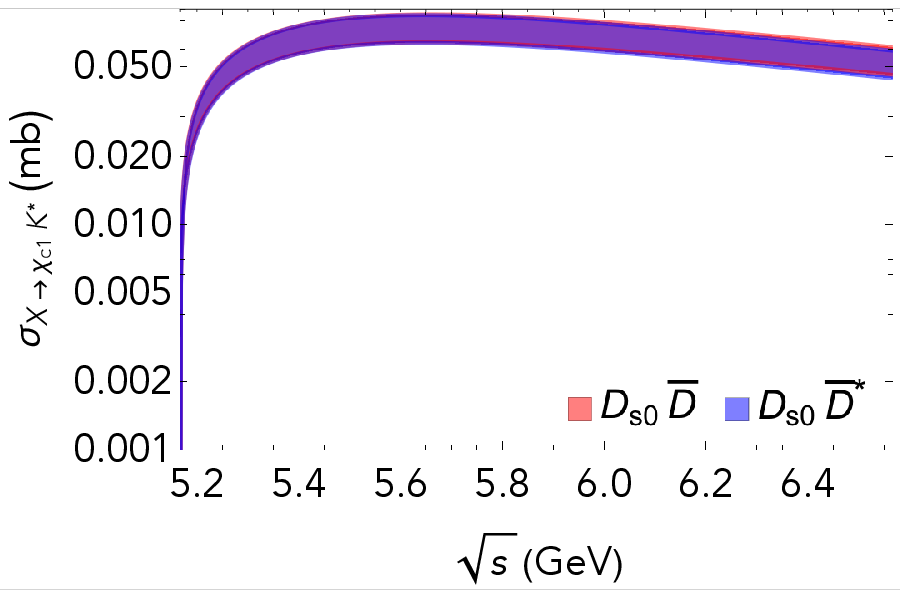}
    \includegraphics[{width=0.32\linewidth}]{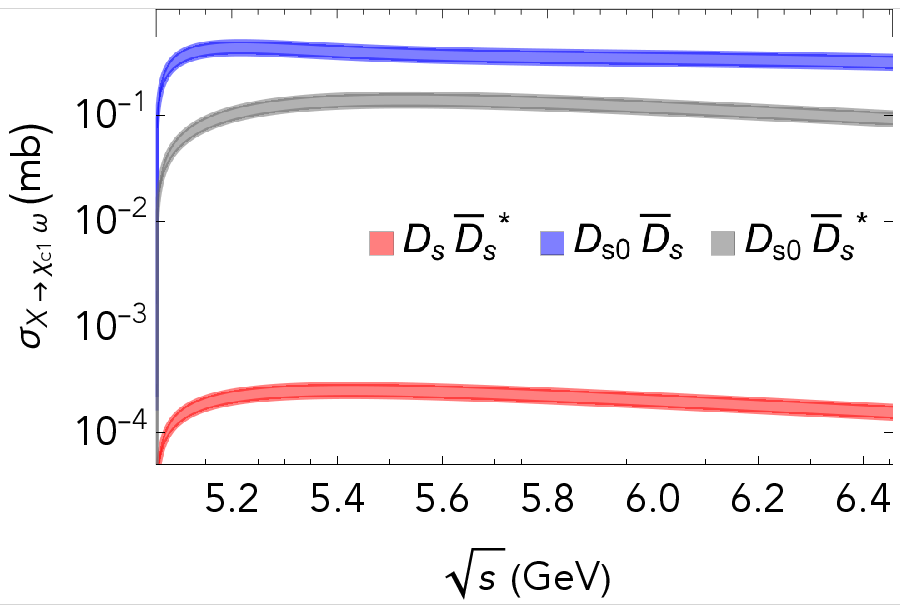}
 \\
    \includegraphics[{width=0.32\linewidth}]{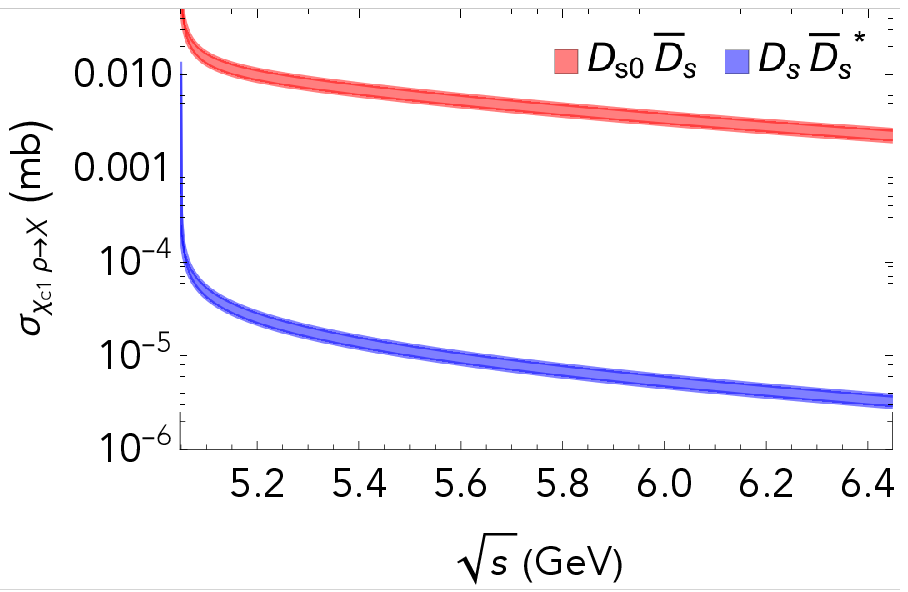}
    \includegraphics[{width=0.32\linewidth}]{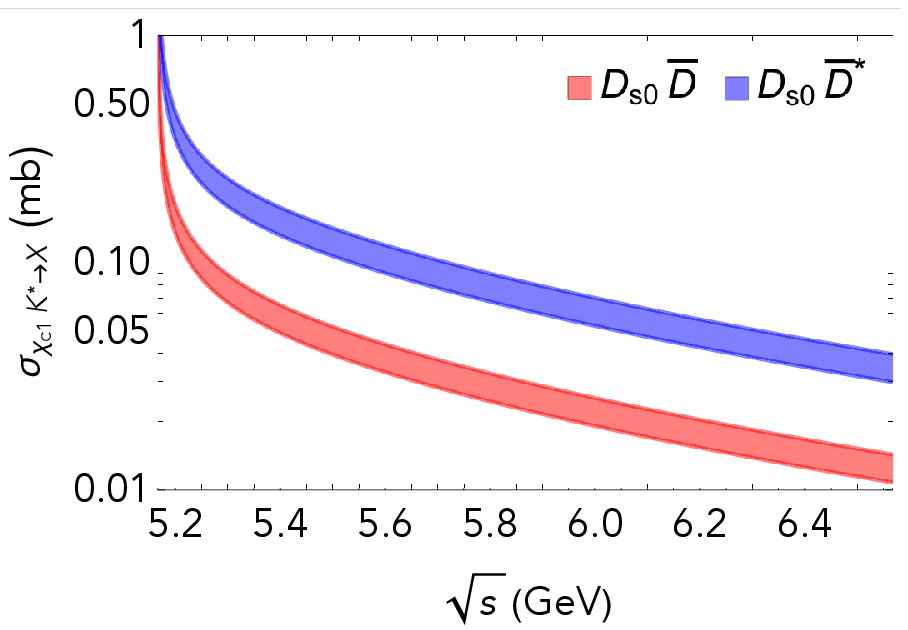}
    \includegraphics[{width=0.32\linewidth}]{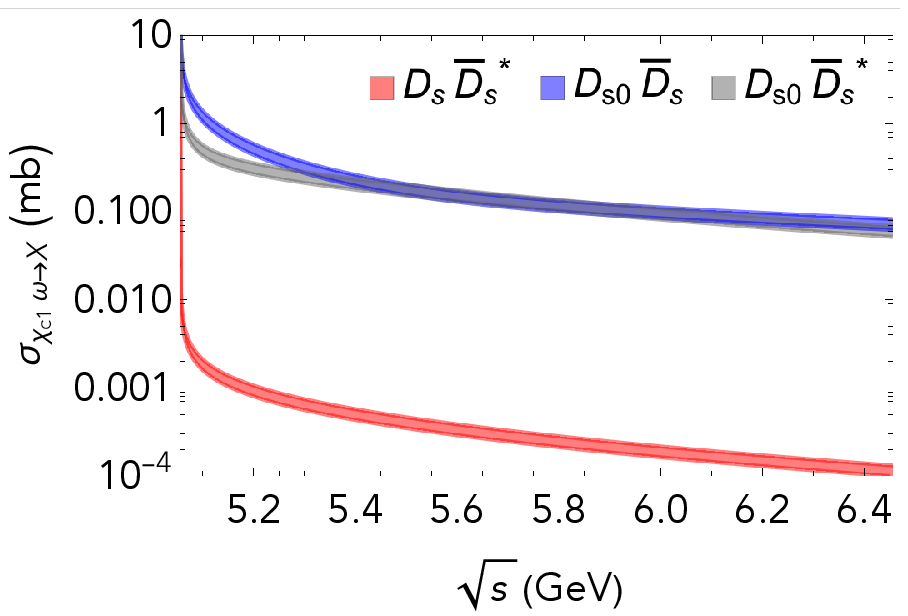}
\caption{Top: cross sections for the production processes $ \chi_{c1}(4274) \rho $
(left),  $ \chi_{c1}(4274) K^* $ (center) and 
$ \chi_{c1}(4274) \omega$ (right), as functions of     
  $\sqrt{s}$. Bottom: cross sections for the corresponding inverse reactions. 
In the process $X\rightarrow \chi_{c1} K^*$ the curves are slightly overlapped.}
\label{Yvector}
\label{suppresion}
\end{figure}
\end{widetext}
\twocolumngrid

\section{Thermal cross sections}

\begin{widetext}
\begin{figure}[h]
\centering
    \includegraphics[{width=0.32\linewidth}]{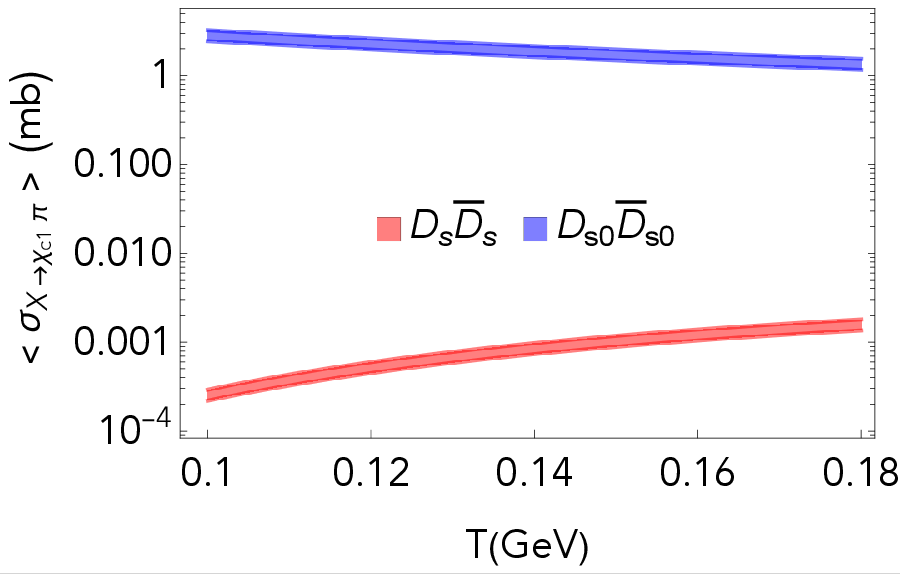}
~  
    \includegraphics[{width=0.32\linewidth}]{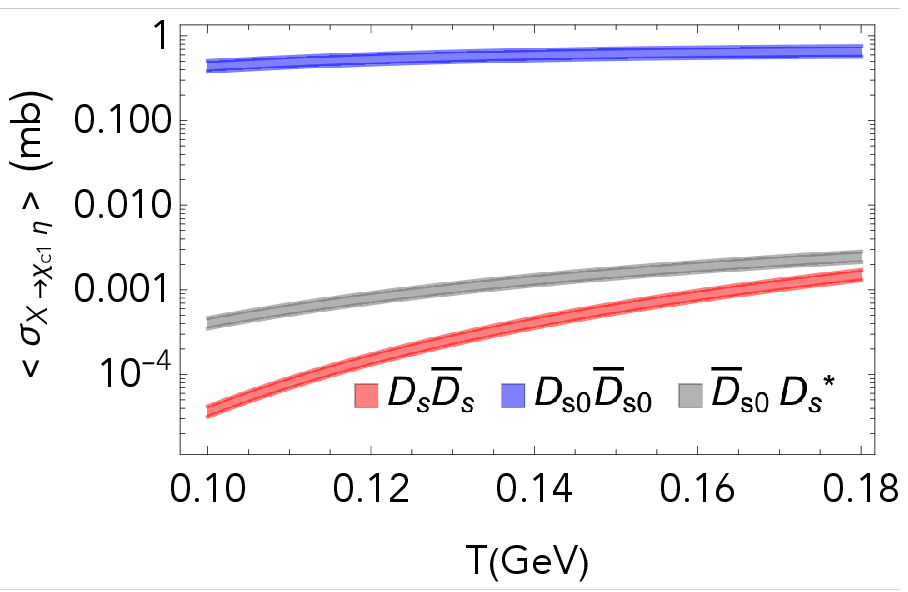}
    \includegraphics[{width=0.32\linewidth}]{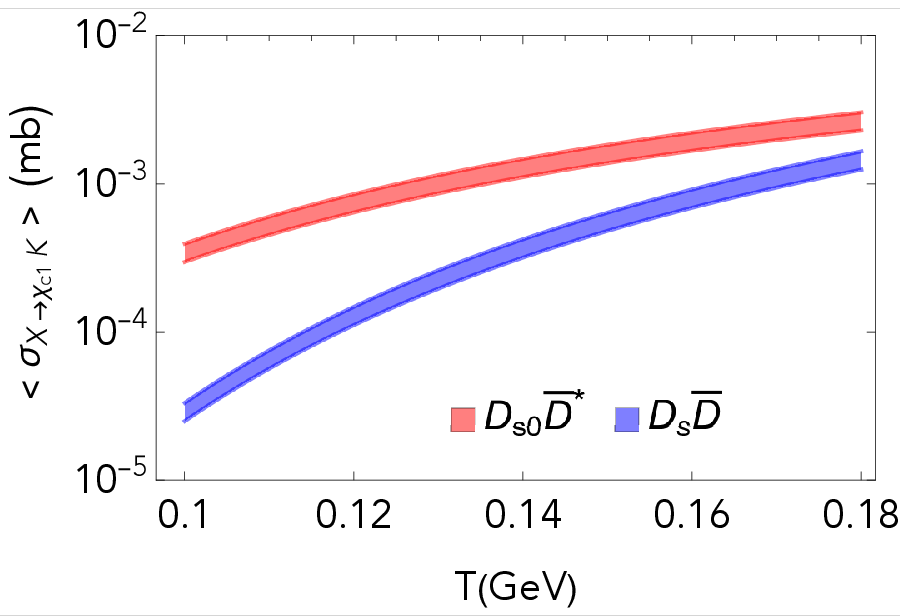}
~
\\
    \includegraphics[{width=0.32\linewidth}]{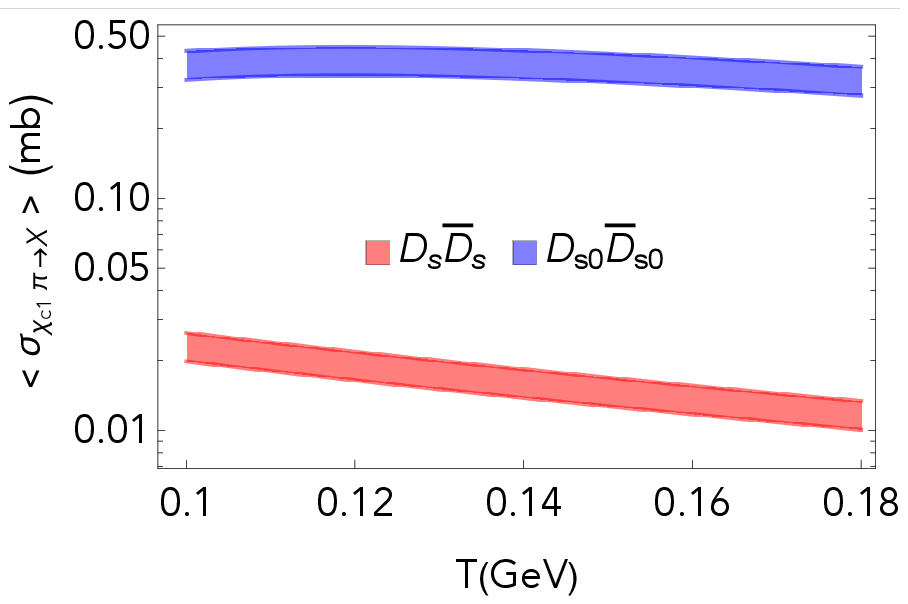}
~ 
    \includegraphics[{width=0.32\linewidth}]{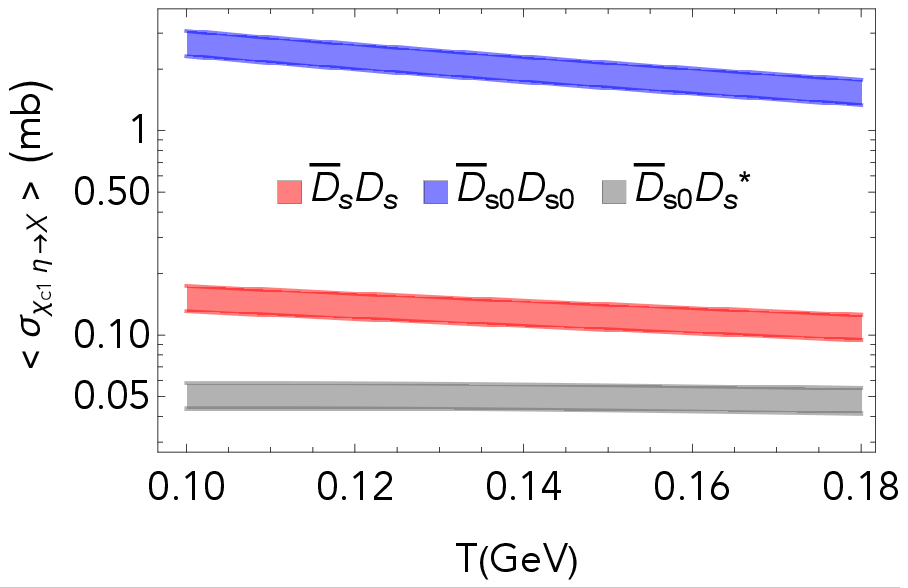}
~ 
    \includegraphics[{width=0.32\linewidth}]{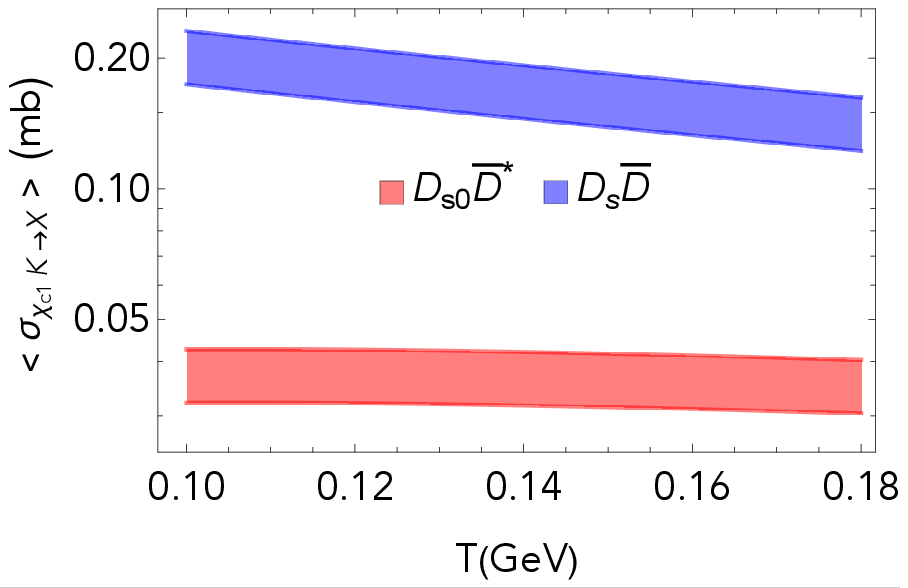}
~ 
\caption{Top: thermal  cross sections for the production processes 
$ \chi_{c1}(4274) \pi  $ (left),
  $ \chi_{c1}(4274) \eta$ (center) and $ \chi_{c1}(4274) K$ (right), as functions of temperature. Bottom: cross sections for the corresponding inverse reactions.}
\label{AvgCSpseudo}
\end{figure}
\end{widetext}

The findings of the previous sections allow us to go ahead and use  
them as input in the analysis of the $\chi_{c1}(4274)$ production 
and suppression        
in a heavy ion collision environment, in which the medium effects become relevant.  
The collision energy is related to the temperature of the hadronic medium, and hence 
we need to evaluate the thermally averaged cross-sections, which are
defined as the cross-sections averaged over the thermal distributions of the 
particles participating in the reactions. For the process $ab\rightarrow cd$ 
they are given by the convolution of vacuum cross-sections and the momentum 
distributions: 
\begin{align}
\left\langle\sigma_{a b \rightarrow c d} v_{a b}\right\rangle=&           
\frac{\int d^3 \mathbf{p}_a d^3 \mathbf{p}_b f_a\left(\mathbf{p}_a\right)   
f_b\left(\mathbf{p}_b\right) \sigma_{a b \rightarrow c d} v_{a b}}{\int d^3  
\mathbf{p}_a d^3 \mathbf{p}_b f_a\left(\mathbf{p}_a\right) 
f_b\left(\mathbf{p}_b\right)} 
\notag\\
=& \frac{1}{4 \beta_a^2 K_2\left(\beta_a\right) \beta_b^2 K_2\left(\beta_b\right)} 
\notag\\
& \times \int_{z_0}^{\infty} d z K_1(z) \sigma\left(s=z^2 T^2\right) 
\notag\\
& \times\left[z^2-\left(\beta_a+\beta_b\right)^2\right]
\left[z^2-\left(\beta_a-\beta_b\right)^2\right],
\end{align}
where $f(p)$ is the Bose-Einstein distribution, $v_{ab}$ is the relative velocity  
of the two initial particles $a$ and $b$; $\beta_i=m_i/T$, where $T$ is the      
temperature; $z_0=max(\beta_a+\beta_b,\beta_c+\beta_d)$, and $K_1$ and $K_2$ are 
the modified Bessel functions of second kind.
In Figs. \ref{AvgCSpseudo} and \ref{AvgCSvector} we plot the thermal cross section 
as functions of temperature. The suppression processes have a weaker dependence  
with the temperature than those for $\chi_{c1}(4274)$ production. 

\onecolumngrid

\begin{widetext}

\onecolumngrid

\begin{figure}[h]
    \includegraphics[{width=0.32\linewidth}]{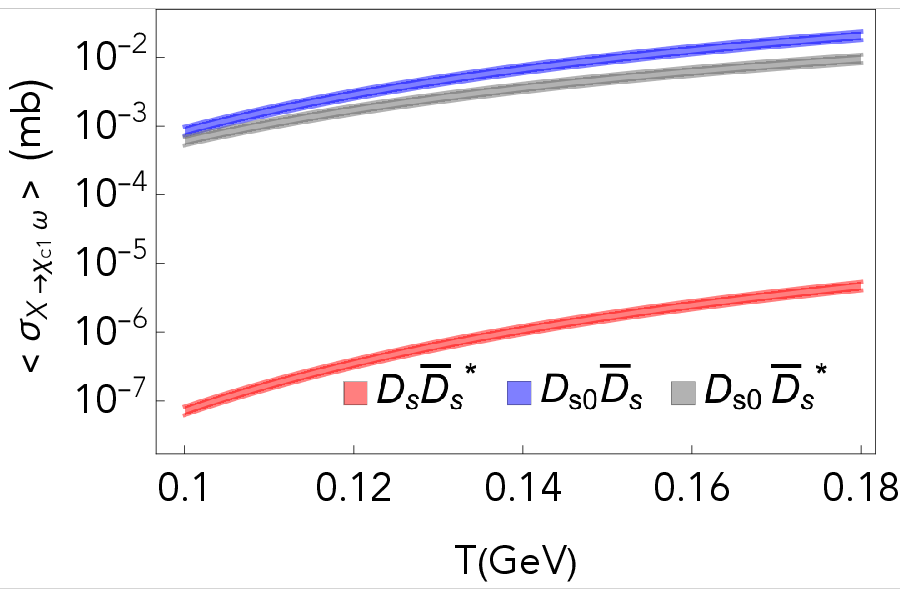}
~ 
    \includegraphics[{width=0.32\linewidth}]{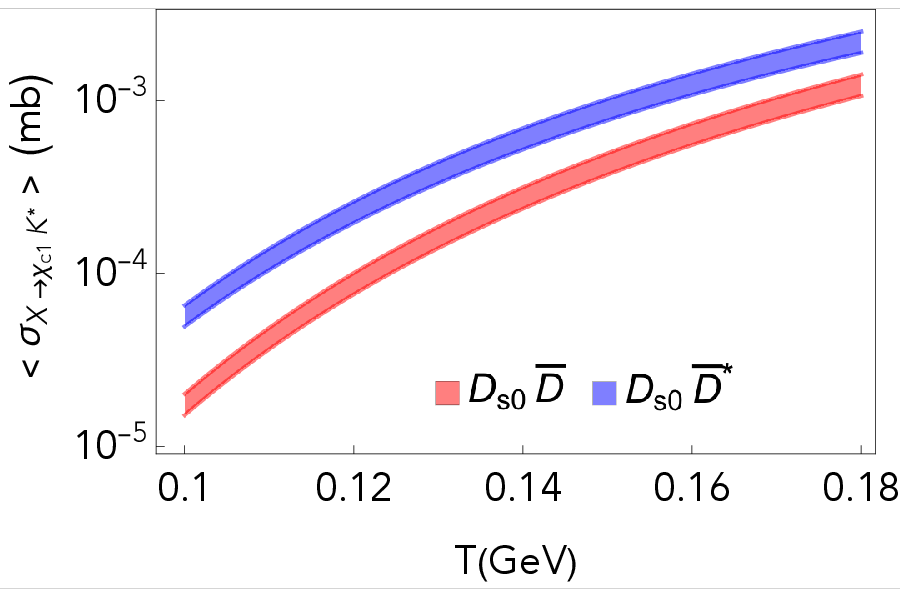}
    \includegraphics[{width=0.32\linewidth}]{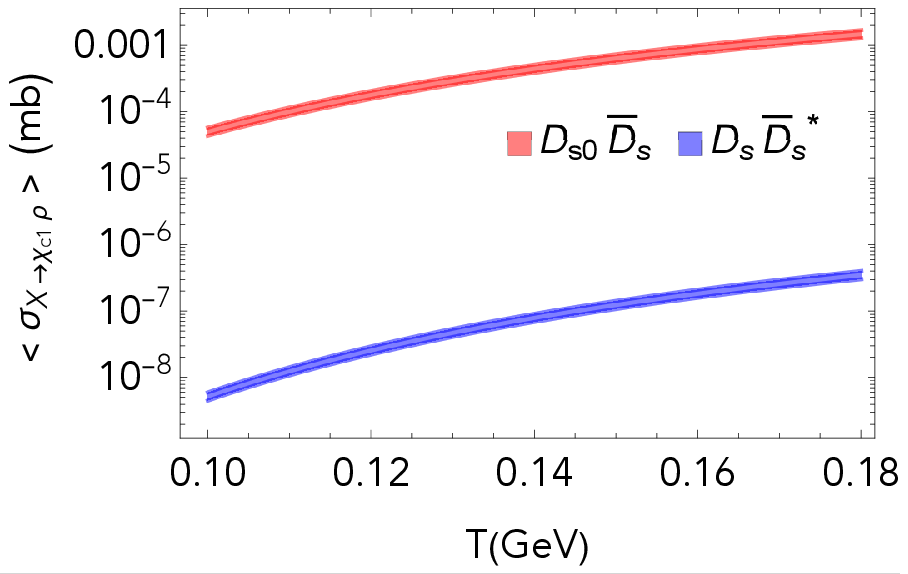}
~ 
\\
    \includegraphics[{width=0.32\linewidth}]{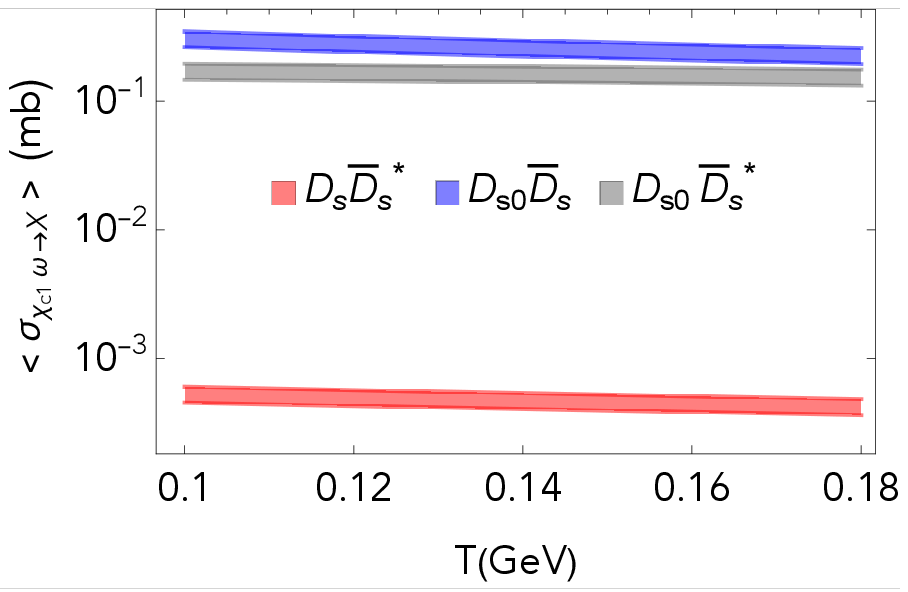}
~ 
    \includegraphics[{width=0.32\linewidth}]{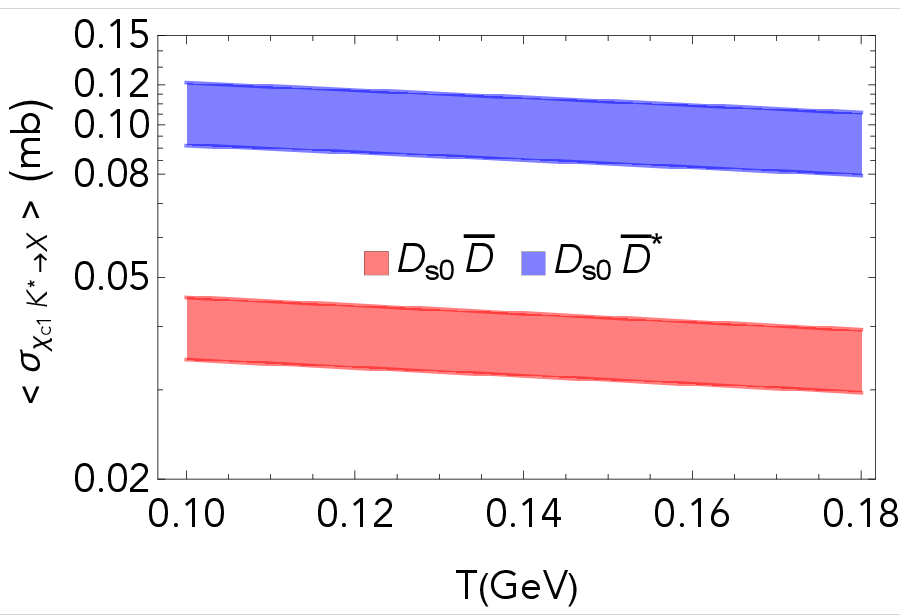}
      \label{rhoAllAvCrSec}
~ 
    \includegraphics[{width=0.32\linewidth}]{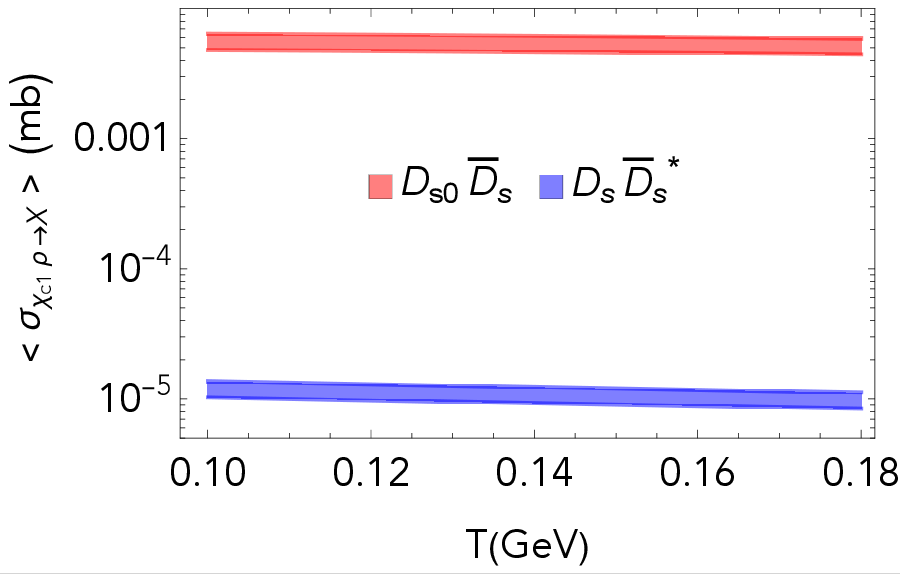}
\caption{Top: thermal  cross sections for the production processes 
$ \chi_{c1}(4274) \rho  $ (left), $ \chi_{c1}(4274) K^* $ (center) and $ \chi_{c1}(4274) \omega$ (right), as   functions of temperature. Bottom: cross sections for the corresponding inverse reactions.}
\label{AvgCSvector}
\end{figure}
\end{widetext}

Comparing all the cross sections shown in the figures, the striking conclusion is that
(unlike the case of most of the other exotic states) the most important process is
$\chi_{c1}(4274)$ production through the reaction 
$D_{s0} \, \bar{D}_{s0} \to \chi_{c1}(4274)\, \pi$. The dominant  absorption reaction 
is $\chi_{c1}(4274) \, \eta \to D_{s0} \, \bar{D}_{s0}$. 
In a hot hadron gas the abundance of
$\eta$ is  larger than the abundances of $D_{s0}$ and $\bar{D}_{s0}$. This may 
compensate the difference in the cross sections, but at this point we can not say 
that there will be a $\chi_{c1}(4274)$ suppression due to hadronic medium effects. 
To know what really happens we must solve the rate equations with the above cross 
sections. This will be addressed in a forthcoming publication.


%

%

%

%

%

%

%



%

%

\section{Conclusions}

\label{Conclusions}

In this work we have studied the interactions of the $\chi_{c1}(4274)$ state with  
light mesons, which are the most abundant particles in the hot hadron gas 
formed in the late stage of heavy ion collisions. Using an effective Lagrangian  
approach, we computed the vacuum and thermal cross sections of $\chi_{c1}(4274)$  
production (accompanied by  light pseudoscalar and vector mesons) and the 
corresponding inverse processes. The coupling constants involving the  
$D_{s0}$ meson were calculated through the VMD model. The results show that the
thermal cross-sections are sizeable. In almost all the cases, the absorption 
cross sections are  larger  than the production ones. However, the largest cross 
section is for $\chi_{c1}(4274)$ production through the reaction
$D_{s0} \, \bar{D}_{s0} \to \chi_{c1}(4274)\, \pi$. Our study strongly motivates 
the use of the obtained cross sections as input to the rate equations, which yield 
the $\chi_{c1}(4274)$ multiplicity during the time evolution of a hot hadron gas. 
Work along this line is in progress.



\begin{acknowledgements}

A.L.M Britto would like to thank H. P. L. Vieira for discussions. This work 
was partly supported  by the Brazilian agencies Conselho Nacional   
de Desenvolvimento Cient\'ifico e Tecnol\'ogico (CNPq) under contracts   
309950/2020-1 (L.M.A.), 400215/2022-5 (L.M.A.), 200567/2022-5 (L.M.A.)),    
and CNPq/FAPERJ under the Project INCT-Física Nuclear e Aplicações (Contract No. 464898/2014-5).


\end{acknowledgements}


\appendix

\section{Amplitudes}

\label{appendice}


The explicit expressions of the amplitudes for the processes represented in Figs. \ref{DIAG1} are:

\begin{eqnarray}
     M^{(a)} &=&
    -\frac{g_{\pi^0D_sD_{s0}}g_{\chi_{c1}D_s\bar{D}_{s0}}}{\sqrt{2}}\epsilon^\mu(p_3)(p_3-2p_1)_\mu\frac{1}{t-m_{D_{s0}}^2},
    \notag\\
    M^{(b)}&=&
    \frac{g_{\pi^0D_sD_{s0}}g_{\chi_{c1}D_s\bar{D}_{s0}}}{\sqrt{2}}\epsilon^\mu(p_3)(p_3-2p_2)_\mu\frac{1}{u-m_{D_{s0}}^2},
    \notag\\
     M^{(c)}&=&
    -\frac{g_{\pi^0 D_sD_{s0}}g_{\chi_{c1}D_s\bar{D}_{s0}}}{\sqrt{2}}\epsilon^\mu(p_3)(p_3-2p_1)_\mu\frac{1}{t-m_{D_s}^2},
    \notag\\
    M^{(d)}&=&
    \frac{g_{\pi^0 D_sD_{s0}}g_{\chi_{c1}D_s\bar{D}_{s0}}}{\sqrt{2}}\epsilon^\mu(p_3)(p_3-2p_2)_\mu\frac{1}{u-m_{D_s}^2},
    \notag
\end{eqnarray}
\begin{eqnarray}
     M^{(e)}&=&
    -\frac{g_{\eta D_sD_{s0}}g_{\chi_{c1}D_s\bar{D}_{s0}}}{\sqrt{2}}\epsilon^\mu(p_3)(p_3-2p_1)_\mu\frac{1}{t-m_{D_{s0}}^2},
    \notag\\
    M^{(f)}&=&
    \frac{g_{\eta D_sD_{s0}}g_{\chi_{c1}D_s\bar{D}_{s0}}}{\sqrt{2}}\epsilon^\mu(p_3)(p_3-2p_2)_\mu\frac{1}{u-m_{D_{s0}}^2},
    \notag\\
     M^{(g)}&=&
    -\frac{g_{\eta D_sD_{s0}}g_{\chi_{c1}D_s\bar{D}_{s0}}}{\sqrt{2}}\epsilon^\mu(p_3)(p_3-2p_1)_\mu\frac{1}{t-m_{D_s}^2} , \notag\\
    M^{(h)}&=&
    \frac{g_{\eta D_sD_{s0}}g_{\chi_{c1}D_s\bar{D}_{s0}}}{\sqrt{2}}\epsilon^\mu(p_3)(p_3-2p_2)_\mu\frac{1}{u-m_{D_s}^2},
    \notag\\
    M^{(i)}&=& \frac{g_{PPV} g_{\chi_{c1}D_s\bar{D}_{s0}}}{\sqrt{3}}
\epsilon^*_\mu(p_2)\epsilon_\nu(p_3)
\notag\\ & & \times
(p_2-2p_4)^\mu(p_3-2p_1)^\nu \frac{1}{t-m_{D_{s0}}^2} , \notag
    \notag \\
M^{(j)}&=&\frac{g_{PPV}g_{\chi_{c1}D_s\bar{D}_{s0}} }{\sqrt{2}}
\epsilon_{\mu}^*(p_2)\epsilon_{\nu}(p_3)\notag\\ 
 & & \times
(-p_2+2p_4)^\mu(2p_1-p_3)^\nu\frac{1}{t-m_{D_s}^2}, \notag\\
M^{(k)}&=&
-\frac{g_{KDD_{s0}}g_{\chi_{c1}D_s\bar{D}_{s0}}}{2}\epsilon^\mu(p_3)
(-p_3+2p_1)_\mu 
\notag\\ & & \times
\frac{1}{t-m_{D_{s0}}^2},
\notag 
\end{eqnarray}
and for those depicted in Fig.~\ref{DIAG2} are
\begin{eqnarray}
    M^{(l)}&=&
    -\frac{g_{\chi_{c1}D_s\bar{D}_{s0}}  g_{\rho D_{s0}{D}_{s0}}}{\sqrt{2}}\epsilon(p_3)^\nu\epsilon(p_4)^\mu
   \notag\\ & & \times (-2p_1+p_4)_\mu(p_3-2p_2)_\nu\frac{1}{u-m_{D_{s0}}^2}, 
   \notag\\
   M^{(m)} &=&   \frac{g_{\rho D_{s0}D_s^*}g_{\chi_{c1}D_s\bar{D}_{s0}}}{\sqrt{2}}
        (p_{4\mu} \epsilon_{ \nu}^{*}(p_2)-p_{4\nu} \epsilon_{\mu}^*(p_2))\notag\\ 
        & & \times
        (p^\nu_2 \epsilon^\mu(p_4)-p_4^\mu\epsilon^\nu(p_4))       
 (2p_1-p_3)_\rho\epsilon^\rho(p_3) \notag\\ 
        & & \times \frac{1}{t-m_{D_{s0}}^2},  \notag
\end{eqnarray}
\begin{eqnarray}
 M^{(n)} &=&
    \frac{g_{PPV}g_{\chi_{c1}D_s\bar{D}_{s0}}}{\sqrt{2}}\epsilon^\mu(p_4)\epsilon^\nu(p_3)\notag\\ & & \times 
    (p_4-2p_2)_\mu(2p_1-p_3)_\nu\frac{1}{t-m_{D_s}^2}, 
        \notag \\
        M^{(o)} &=&
     \frac{g_{VVP}g_{\chi_{c1}D_s D_{s0}} }{{2}}\epsilon^{\mu\nu\alpha\beta}\varepsilon^\rho(p_3)\varepsilon^\beta(p_4)\varepsilon^\nu(p_2)\notag\\ & & \times
    (p_4)_\alpha (p_2)_\mu(2p_1-p_3)_\rho \frac{1}{t-m_{D_s}^2}
\notag\\
M^{(p)}&=& 
    -
    \frac{ g_{PPV}g_{\chi_{c1}D_s\bar{D}_{s0}} }{\sqrt{3}}\epsilon(p_3)^\nu\epsilon(p_4)^\mu \notag\\ 
    & & \times
    (2p_2-p_4)_\mu(2p_1-p_3)_\nu\frac{1}{t-m_{D_s}^2}
    \notag\\
    M^{(q)}
      &=&   
    \frac{g_{\chi_{c1}D_s\bar{D}_{s0}}  g_{\omega D_ {s0}D_s}}{\sqrt{2}}\epsilon(p_3)^\nu\epsilon(p_4)^\mu \notag\\ 
    & & \times
    (p_4-2p_1)_\mu(p_3-2p_2)_\nu\frac{1}{u-m_{D_{s0}}^2}
    \notag\\
    M^{(r)}
       &=&
      \frac{g_{\omega D_{s0}D_s^*}g_{\chi_{c1}D_s\bar{D}_{s0}}}{\sqrt{2}}
        (p_{4\mu} \epsilon_{ \nu}^{*}(p_2)-p_{4\nu} \epsilon_{\mu}^*(p_2))\notag\\ 
        & & \times
        (p^\nu_2 \epsilon^\mu(p_4)-p_4^\mu\epsilon^\nu(p_4))
        [ (2p_1-p_3)_\rho]\epsilon^\rho(p_3)
        \notag\\ 
        & & \times \frac{1}{t-m_{D_{s0}}^2}
        \notag \\
        M^{(s)}
       &=&
       \frac{g_{VVP}g_{\chi_{c1}D_s\bar{D}_{s0}}}{\sqrt{6}}
     \epsilon^{\mu\nu\alpha\beta}
\varepsilon^\rho(p_3)\varepsilon^\nu(p_4)\varepsilon^{*\beta}(p_2)\notag\\ 
& & \times(p_4)_\mu (p_2)_\alpha 
     (-2p_1+p_3)_\rho
     \frac{1}{t-m_{Ds}^2}\notag
\end{eqnarray}

In these expressions $p_1$ and $p_2$ are the momenta of the initial states and $p_3$ and $p_4$ are the momenta of the final states. The $\varepsilon^\mu(p_i)$ is the polarization of vector states with momenta $p_i$; $u$ and $t$ are the Mandelstam variables, which jointly with $s$, are defined as follows: $s=(p_1+p_2)^2$, $u=(p_1-p_4)^2$ and $t=(p_1-p_3)^2$.


\end{document}